\documentclass[12pt,english]{article}

\usepackage[T1]{fontenc}
\usepackage[utf8]{inputenc} 
\usepackage[scaled=0.95]{helvet} 

\usepackage{setspace}
\usepackage{ragged2e}

\newenvironment{HarvardBibliography}{%
  \par\begingroup
  \setstretch{1.5}%
  \setlength{\parindent}{0pt}%
  \setlength{\parskip}{0.55\baselineskip}%
  \everypar{\hangindent=1.27cm\hangafter=1}%
  \RaggedRight
}{%
  \par\endgroup
}

\usepackage{geometry}
\usepackage[titletoc,title]{appendix}

\raggedright

\usepackage{sectsty}
\allsectionsfont{\normalsize\raggedright\centering}

\usepackage[]{tocloft}
\addtocontents{toc}{\cftpagenumbersoff{section}}
\addtocontents{toc}{\cftpagenumbersoff{subsection}}

\makeatletter
\renewcommand{\@cftmaketoctitle}{}
\makeatother

\usepackage[]{footmisc}

\usepackage{authblk}

\usepackage{graphicx}
\usepackage{babel}

\usepackage{amsmath}
\usepackage{amsfonts}

\usepackage{framed}

\usepackage{xifthen}


\numberwithin{equation}{section}

\usepackage{txfonts}
\usepackage[T1]{fontenc}

\newcommand{\citetbjps}[2][]{\ifthenelse{\equal{#1}{}}{\citeauthor{#2} ([\citeyear{#2}])}{\citeauthor{#2} ([\citeyear{#2}], #1)}}
\newcommand{\citealtbjps}[2][]{\ifthenelse{\equal{#1}{}}{\citeauthor{#2} [\citeyear{#2}]}{\citeauthor{#2} [\citeyear{#2}], #1}}
\newcommand{\citepbjps}[2][]{\ifthenelse{\equal{#1}{}}{(\citeauthor{#2} [\citeyear{#2}])}{(\citeauthor{#2} [\citeyear{#2}], #1)}}
\newcommand{\citeyearbjps}[2][]{\ifthenelse{\equal{#1}{}}{[\citeyear{#2}]}{[\citeyear{#2}], #1}}
\newcommand{\citeyearparbjps}[2][]{\ifthenelse{\equal{#1}{}}{([\citeyear{#2}])}{([\citeyear{#2}], #1)}}
\newcommand{\citeposbjps}[2][]{\ifthenelse{\equal{#1}{}}{\citeauthor{#2}'s ([\citeyear{#2}])}{\citeauthor{#2}'s ([\citeyear{#2}], #1)}}

\begin{document}

\title{ \textbf{What is Physics For? \\ {\large Why Classical Physics is not the Limit of Quantum Mechanics}} }

\author[1]{{Adam Frank }}
\affil[1]{{\normalsize Department of Physics and Astronomy, University of Rochester, Rochester, NY, USA}\vspace{-2mm}} 
\author[2]{{Jacques L. Pienaar}}
\affil[2]{{\normalsize Instituto de Física, Universidade Federal do Rio de Janeiro, RJ. Brazil}\vspace{-2mm}}
\author[3]{{Michel Bitbol}}
\affil[3]{{\normalsize Archives Husserl, École Normale Supérieure / CNRS, Paris, France}\vspace{-2mm}}
\author[4]{{Harald Wiltsche}}
\affil[4]{{\normalsize Division of Philosophy and Applied Ethics, Linköping University, Sweden}\vspace{-2mm}}
\author[2]{{Gabriela~Barreto~Lemos}}
\affil[2]{{\normalsize Instituto de Física, Universidade Federal do Rio de Janeiro, RJ. Brazil}\vspace{-2mm}}
\author[5]{{Marcelo Gleiser}}
\affil[5]{{\normalsize Department of Physics and Astronomy, Dartmouth College, Hanover, NH 03755, USA}\vspace{-2mm}}
\author[6]{{Marcus Appleby}}
\affil[6]{{\normalsize School of Physics, University of Sydney, Sydney, NSW 2006, Australia}}

\date{}

\maketitle

\thispagestyle{empty}

\begin{abstract}
Quantum mechanics is justifiably held to be the most successful physical theory ever created; yet no universally accepted interpretation of its mathematical formalism exists after a century of debate. We examine the specific claim that classical physics stands as the low-action limit of quantum mechanics, arguing that the deep interpretive difficulties exposed by the quantum formalism—the measurement problem, the subject–object divide, and the nature of scientific representation—were already latent in classical physics, constituting what we call a blind spot of objectivity. Taking quantum theory as an invitation to re-examine the implicit metaphysical commitments bundled with 'classical physics', we explore two key issues—the objective–subjective split and the nature of theoretical representation—and argue that quantum mechanics does not represent a break with classical physics per se but rather brings previously unacknowledged philosophical assumptions into the limelight. We discuss the London–Bauer interpretation and QBism as exemplars of interpretations that embrace this broader reconceptualisation.
\end{abstract}

\section{Introduction}

Quantum Mechanics is justifiably claimed to be the most successful and most accurate scientific theory ever created. In spite of this success, there remains no universally accepted interpretation for the mathematical formalism on which quantum mechanics is based (Cabello, 2017). The fact that, even a century since its founding, this bedrock of physical theory is seen as needing an interpretation is the subject of this paper. As we will see, the question highlights a far deeper and broader set of issues that pivot about our expectations for the relationship between scientific theories and the reality those theories are meant to describe.  

Our goal in this work is to address the specific claim that classical physics can be seen as the limit of quantum mechanics. The idea of one theory being the limit of another more fundamental one occurs in many domains of the physical sciences (Palacios, 2024).  Newtonian mechanics, for example, is seen as being the low velocity limit for Special Relativity which includes all velocities up to those approaching the speed of light.  Likewise, Newtonian gravity is the low-curvature limit of a background spacetime manifold.  General Relativity is the more complete theory required when manifold curvatures are high as the ratio of mass and radius for an object ($M/R$) approaches the ratio $c^2/G$, where c is the speed of light and $G$ is Newton' s constant. 

The relationship between classical and quantum physics can in many cases also be framed as one of limits.  A well-known example is the Ehrenfest Theorem, which states that the expectation values of the quantum position and momentum operators obey Newton’s classical equations of motion (Ehrenfest, 1927). More generally, the Correspondence Principle states that classical theory applies in the limit of large quantum numbers, such that for large orbits and large energies quantum physics must agree with classical physics (Bohr, 1920). For example, classical theory applies when the characteristic action (energy / time) of a system is large compared to Planck’s constant $h$.  These examples seem to suggest that Quantum Mechanics can be regarded as the more general theory which holds even when the action is less than or comparable to $h$, supporting the idea that classical mechanics is the low action limit of quantum mechanics. Max Planck was the first to illustrate this idea in a specific example, when he showed that his blackbody radiation law reproduces the classical Rayleigh-Jeans law, which is only valid for large wavelengths (Planck, 1906).

While this perspective is no doubt useful, it is noteworthy that unlike the preceding examples from classical physics, there is no universal method for taking a classical limit in quantum theory. When it can be done, the method is usually domain-specific; moreover there are examples of quantum models that have no well-defined classical limit (e.g. Pineda and Prosen, 2007). The complications that beset all attempts to treat the Classical / Quantum relation as one of limits points to an additional difficulty  often ignored, namely, the question of how these two theories should be interpreted. We now turn to this pressing issue.

For the majority of working physicists there is no question as to the interpretation of classical physics. Unless one is specifically interested in the philosophy of science, classical physics appears to speak for itself. The mathematical machinery describes “objects” (i.e. particles, rods, springs) that can unproblematically be visualized from an imagined, objective third-person perspective.  Going further, in classical physics there was no need to ask about the object’s properties (position, momentum etc) existing before they were observed. Nothing prevents one from assuming that things existed “out there” ontologically, with exactly the properties the theory specified.

Nevertheless, a more careful historical study of the philosophical understanding of classical mechanics reveals that the theory was originally far from being considered conceptually clear and straightforward. Many concerns were raised about it, some of which we shall explore in what follows. For example: 1) Why did classical physics content itself with mathematical laws connecting quantitative variables instead of identifying the “cause” of gravitation or of electromagnetic interactions? 2) Why is motion considered relative instead of absolute in some cases (inertial frames) and not in others (accelerated frames)? In addition, several interpretations or reconstructions of classical mechanics had been proposed. For example, the standard interpretation in terms of discrete material bodies endowed with kinematic properties was challenged by a non-standard interpretation in terms of a continuum of energy whose intense localizations may mimic the effect of material bodies. More radically, Mach argued against interpreting mechanics in terms of hypothetical microscopic models, insisting instead that physics be formulated strictly in terms of observable quantities (Mach, 1919). Boltzmann, meanwhile, made a spirited defense of microscopic models of mechanics (Boltzmann, 1979). 

These debates notwithstanding, in the late 19th century it was generally accepted that nothing \textit{in principle} precluded the explanation of observed physical phenomena by an underlying mechanistic model (Stern, 1998), and so the door remained open to interpreting the physicists’ models as approximations to an ultimate description of microscopic reality, such as would be knowable to a hypothetical knower (e.g. God) who would be free from all the impediments that presumably obstructed the inquiry of such fallible, finite beings as we are. To support or discredit this idea, one could only draw upon philosophical arguments, not physical facts; for the observations made in laboratories were of a neutral character, neither encouraging nor challenging the hopes of those who sought a detailed and complete picture of reality itself.  As we will see, this was never the case for quantum mechanics.

For the working physicist, the formalism of quantum mechanics does not offer the same kinds of imaginative guides as to the reality of its own mathematically described objects as does classical physics.  An electron, for example, cannot be modeled strictly as a wave or a particle — that is, it does not possess explicitly wave-like or particle-like properties — until a measurement is made. It can also exist in so-called “superposition” states, which do not correspond to a definite value for any single property, yet also cannot be straightforwardly understood as states where the electron simultaneously possesses multiple definite values of a given property (such as its spin angular momentum), since only one of the possible values is realized when the property in question is measured. For those trained first in classical physics (which is every physics student) this shift from the classical to the quantum can be profoundly disconcerting.  As Niels Bohr, one of quantum mechanics’ founders put it, “For those who are not shocked when they first come across quantum theory cannot possibly have understood it.” (Niels Bohr, as quoted in Heisenberg, 1971)  

Thus while classical mechanics seemed to implicitly accommodate realist interpretations, quantum mechanics forced its users to explicitly justify their choice of interpretation, particularly if it was a realist one. This development occasioned a remarkable intrusion of philosophical discourse into the world of the working physicist, leading some to rebel against the trend by doubling down on a pragmatic stance that ignored interpretive questions, a position caricatured by physicist David Mermin as “shut up and calculate” (Mermin, 1989). Of course when Mermin wrote of the “Shut up and calculate” attitude he was doing so ironically,  asking how any physicist would be so sanguine as to simply use the quantum formalism while ignoring its glaring provocation to those who sought a microscopic model of reality. One could still turn one’s back on the problem, certainly, but one could no longer avoid admitting that there was a problem. Refusing to take a stand on questions of interpretation thus became an explicit choice; a deliberate, willful decision that classical physics never demanded. 

Returning now to the question of limits, this inability to simply imagine and interpret the contents of the quantum formalism is why the question of classical physics being a limit of quantum mechanics is not as straight-forward as it is in other domains.  As we will discuss below, the central difficulty in interpreting quantum mechanics lies in the so-called “measurement problem”. The essential role of measurement in the theory infects the question of limits. The difficulty stems from the fact that although a measuring device is itself made of atoms, the results of a measurement must be intelligible to a macroscopic observer, ultimately a human. In practice, this is achieved through a series of signal amplifications that establish a bridge between the (sensorially inaccessible) quantum realm and the (sensorially accessible) classical world. Where the boundaries between the quantum system and the measuring apparatus should be placed, and how to describe the interaction across that boundary, is treated differently in different interpretations (Landsman 2007).  In this way, the question of interpretation makes a direct appearance in questions about limits and the “real” nature of the systems on both sides of the divide which remain contested.

In this paper we wish to examine the issue of classical physics as a limit of quantum mechanics in a specific framework. Our contention is that the deep and vexing questions raised by the quantum formalism were always present in classical physics.  Issues such as the pre-existing reality of unobserved system properties and the centrality of measurement (and perhaps measurers i.e. observers) in manifesting system properties had simply been cast into a “blind spot” and so for the most part went unseen and unrecognized\footnote{As noted earlier, philosophers and physicists including Mach, Boltzmann, and others, did openly debate these issues; however prior to quantum theory most physicists could consistently ignore such issues, as they were largely independent of the observable predictions; this changed with quantum theory.
}.  In what follows we will argue that quantum mechanics simply forced these questions into the limelight, ushering in a new era of what philosopher-physicist Abner Shimony dubbed “experimental metaphysics” (Shimony, 1993).  Quantum mechanics, both in its unexpected predictions for experiments and the unexpected experimental results that precipitated its development (such as black-body radiation and the photoelectric effect), laid bare for the first time a host of previously unacknowledged metaphysical commitments (the aforementioned blind spot), forcing physicists to defend or abandon these commitments in a way that classical physics never required.  

This has bearing on the question of limits: for if we take “classical physics” to include a metaphysical commitment to the idea that theories seek to describe reality “in itself” – as exemplified by the ideal of a hypothetical complete microscopic mechanical model of reality – then to the extent that quantum physics challenges this ideal, the former cannot possibly be regarded as a limit of the latter, as there is an insurmountable metaphysical gap between the two. On the other hand, we can choose to take the metaphysical lessons of quantum theory as an invitation to re-examine the implicit metaphysical commitments that are often uncritically bundled together with the conception of “classical physics”. It is the latter path that we aim to pursue here.

In what follows we explore two key issues in the interpretation of quantum mechanics which, we will show, also occur in the context of classical physics (and, in fact, most of science).  The first is the question of the objective-subjective split in which the systems under study are assumed to be fully independent of those carrying out the study.  The second is the question of representation which asks how “objects” in a theory are held to correspond to reality.

\section{Challenging the objective-subjective divide}

The distinction between subjective and objective knowledge is often claimed to be the essential feature that distinguishes science from other human activities. It was the development of classical physics that firmly established this distinction. From Newton onward, classical physics was held as the premier exemplar for the production of objective knowledge based on the precision of its theoretical predictions and accuracy of the correspondence of those predictions with experimental data. 

In this section we argue that there is an essential blind spot in assumptions about the role of objectivity in physics and in science generally. The problem with assuming that physics provides an objective view of a world independent of observer-participants was revealed with extraordinary clarity by quantum mechanics; we further argue that this difficulty was already looming in classical physics. 

What we will call the blind spot of objectivity has two interrelated aspects: (i) the unquestioned assumption that objects or “systems” pre-exist our activity of accessing them (by detection or measurement); (ii) the associated attempt to ignore or completely subtract the subjective components of such activity (lived experience, embodied technological achievements, projects of knowledge, etc.) from the final outcome of our theoretical and experimental endeavors. 

This blind spot has not hitherto passed unnoticed or unchallenged. Contemporary feminist epistemologies have argued that knowledge is always produced by situated knowers, and that acknowledging this situatedness is not a concession to relativism but a condition for achieving more reliable forms of objectivity (Harding, 1995; Toole, 2022). In this view, perspective is a feature of inquiry which needs to be made explicit and available to critical scrutiny, instead of being swept under the rug.

Insofar as physicists prior to quantum theory implicitly and uncritically presupposed the assumptions above, they exhibited a blind spot, as they overlooked the necessary role played by the perspectival experiences of situated observers in doing science and achieving the ideal of objectivity\footnote{With some notable exceptions; for instance, Mach believed that epistemological issues were directly relevant to practical physics, and his view on this point also influenced Einstein and Heisenberg.}. At the same time, adopting these assumptions left both practitioners and philosophers unable to account for the essential role of this very “embodied experience” (as we call it) in enabling physicists to carry out their work (Bitbol, 2002, Frank et al, 2024).

In classical physics the problems created by this blind spot of objectivity were, as we have argued, primarily metatheoretical (epistemological) in character. These problems, raised by the ignorance of the subjective background of classical science, were documented by Kant in his philosophical reflection on Newtonian mechanics. According to Kant, “If nature meant the existence of things \textit{in themselves}, we would never be able to cognize it, either \textit{a priori} or \textit{a posteriori}” (Kant, 2004, p46). Let us unpack this statement further.

On the one hand we are not able to cognize nature \textit{a priori} by imposing on it the structures of our sensibility and understanding. This would mean beginning with our already existing intuitive and conceptual structures — such as space as extension, time as flow, and causality as an ordered sequence of interactions — and hoping that nature itself conforms to them. But Kant argued that there is no reason for nature-in-itself to be  congruent with our intellectual faculties. The only way for there to be an exact isomorphism between the structure of our thinking and the deep and detailed structures underlying physical reality would be if some kind of pre-established harmony existed as Leibniz described in his attempt to account for the apparent causality linking psychophysical monads (i.e. a version of the mind-body problem).

While \textit{a priori} understanding of Nature is impossible without positing a pre-existing harmony, Kant also challenged a complete \textit{a posteriori} understanding of Nature.  The problem in this case is that we have no means of justifying our theories other than by comparing the tentative knowledge they convey with a (utopian) perfect knowledge of nature-in-itself. In Kant’s own words “as the object is out of us and the cognition in us, we never can but judge whether our cognition of the object agrees with the cognition of the object” (Kant, 1819, p64). This argument uses the fact that our senses only give us access to the phenomenal world —the world of appearances as it is structured by our minds. We can never know the “thing-in-itself” (the \textit{noumenal} world) independently of our experience of it. Therefore, any knowledge we gain through experience (\textit{a posteriori}) is necessarily limited and shaped by our own cognitive faculties. 

Let us then accept, as Kant did, that our objects are merely phenomena, byproducts of a creative interaction between us and what we are investigating. And let’s also recognize that the laws our understanding ascribes to the natural behavior of these objects are instantiations of the very rules that allow us to \textit{define} them by extracting invariants from what appears to us. These two reflective claims alone account for the balance of \textit{a priori} and \textit{a posteriori} knowledge, for the rational hypothesizing and experimental testing that characterizes scientific research. Thus, in classical physics, epistemological clarity is achieved only by abandoning the claim of the absolute pre-existence of objects plus their properties, and by making room for the relational and subjective components of knowledge.

Given these questions about the possibilities of objectivity - the assumption that science gives us an impersonal 3rd person “Gods-eye” perspective - let us now consider the appearance of such questions in Quantum mechanics.  We have seen above how, in the context of classical physics, this claim is undermined by considerations of epistemological self-understanding. In quantum physics such claims are challenged by the very intelligibility of its theoretical structure. The persistent attempt to regard the symbols of quantum formalism (such as state vectors) as faithful representations of physical systems has given rise to countless paradoxes and misunderstandings, not the least of which is the “measurement problem” (documented below). Even so, there is a persistent reluctance among quantum physicists to recognize and fully overcome their blind spot. 

Indeed, what we see as a blind spot seems to be just a side effect of a desirable (and even definitional) choice of the scientific ideal they have embraced. This manifests in a choice to eliminate the subjective biases, the sensed qualities, the parochial views, in favor of a struggle to capture the world as it would be in our absence.  This is done by privileging mathematical properties and a “view from nowhere” as philosopher Thomas Nagel put it (Nagel, 1986). 

Yet this ideal of aperspectival objectivity, appealing though it may seem on the surface,  has been subjected to sustained critique. Toole (2022) and others have contested the ideal of a “view from nowhere” arguing that evidence and observation are intelligible only with background assumptions and social practices. Attempts to eliminate all perspective risk stripping away the conceptual, practical, and experimental contexts that allow observations to count as evidence in the first place . In their struggle to find a way around the profound ambiguities concerning subject-participant and object in quantum mechanics, contemporary physicists become heirs to those “men of science” of “the mid-nineteenth century”, who, according to Lorraine Daston and Peter Galison (2007, p34), “began to fret openly about a new kind of obstacle to knowledge: themselves (\textit{qua} subjects).” The scientists’ expectations, theoretical commitments, interpretational habits and other subjective elements began to be seen as a threat to objectivity, as they could mask or distort the observations of nature. In other words, the scientist was seen no longer merely as a bearer of faculties or reason, but as a willful subject whose intervention had to be minimized.  As Daston and Galison have described, a new notion of objectivity emerged in this context as: “the suppression of some aspect of the self, the countering of subjectivity”. 

The resistance to compromise on this ideal (a perfect view from nowhere) - the refusal to let down one’s guard in the face of the threat of a backlash from the subjective -  tended to increase after the 1940s and the 1950s. The philosophical influence of Bohr and Heisenberg, and their insistence on the impossibility of subtracting the effects of measurement from the measured variables prevailed only for two decades after the advent of quantum mechanics. Afterwards, Everett’s \textit{Many Worlds Interpretation} and its variants (reviewed in Vaidman, 2024) and \textit{Objective Collapse} models (reviewed in Ghirardi and Bassi, 2020) attempted to reassert the perfectly objective frame. These were offered as a means around the perspective Bohr and Heisenberg were beginning to consider.  But such a persistent, if not growing, unwillingness to acknowledge the blind spot and understand its consequences was based on a radical misunderstanding of what subjectivity and objectivity mean when it comes to science. 

Let us begin with subjectivity. Two senses of “subjectivity” are to be distinguished. To be “subjective” in the first sense is to make universal claims on the basis of our individual biased point of view. This is obviously to be avoided at all costs in the edification of a science. But the second sense of “subjectivity” is by no means averse to the scientific project. To make room for subjectivity in this second sense just means recognizing the obvious fact that what we try to know (say, the manifest world) is: (a) seen and acted upon from somewhere, and (b) co-constituted by acting upon and perceiving “it”.  In other words, we can not help but push on the world, and the world, in response, cannot help but push back. This is an essential point which unites the epistemological grounding of both classical and quantum physics.  The essential difference is that quantum mechanics made the problem impossible to ignore. 

In this spirit, Sandra Harding’s notion of \textit{strong objectivity} proposes that objectivity is strengthened—not weakened—when the social position and background assumptions of the knower are themselves subjected to systematic scrutiny (Harding, 1995). Strong objectivity requires placing the subject of knowledge on the same critical plane as the object of knowledge, thereby expanding the range of perspectives brought into scientific inquiry (Toole, 2022). The aim is not to privilege any particular standpoint, but to widen the field of critical engagement so that assumptions shared within a community become visible and corrigible (Toole, 2022).

We also note here our specific meaning of “co-constitute” vs the sometimes used “co-create”. While the latter is associated with fabricating, with bringing something that wasn't there before into being, the former has the more nuanced meaning of not conjuring something ex nihilo, but of disclosing or organizing sense within a given horizon of meaning. Hence, while creation implies ontological production, constitution concerns phenomenological clarification: it names the way in which meaning comes to be for a subject without thereby claiming that that subject invents its object. This difference, subtle as it may seem, marks the distance between an idealist fabrication of reality and a transcendental elucidation of how reality is given as such (Wiltsche 2025). 

The inescapable situatedness of knowledge was an issue for philosophers of science and physicists alike, well before quantum mechanics. In physics, it finds its clearest expression in relativity theory, through the need of specifying a coordinate system in order to connect statements of “invariant geometry” (the formal language of 4-vectors, tensors, and manifolds) to observations made by scientists in a specific laboratory setting. A useful analogy may be found in comparing theoretical models to \textit{tools}, rather than to \textit{descriptions}; as with any tool, a model requires “handles” that allow its users to grasp it; in relativity, the “handles” are coordinate systems. Seeing from somewhere is \textit{de facto} integrated into relativity theory through the zero point of coordinate systems; as Hermann Weyl (1949) put it: “the coordinate system remains as the inevitable residue of the annihilation of the ego”. Even the so-called “coordinate free” terms of relativity are inextricable from the groups of transformations that leave them invariant, which in turn depend on what is preserved when we experimentally or intellectually vary the \textit{standpoint} from which our observations are made. 

Such considerations lead us to contemplate more nuanced views of objectivity, such as that of Kant, for whom “the line between the objective and the subjective generally runs between universal and particular” (Daston and Galison, 2007, 30), that is, between what is valid for every rational subject (from anywhere) and what is valid only for one or a few (from a specific place).  In the famous measurement problem - as it was articulated through the work of Von Neumann and later London and Bauer (1939) - Kant’s insight plays an important role in articulating the meaning of objectivity in the new quantum formalism, even though this insight was initially formulated in the context of classical physics.

As for co-constitution, this too was discussed long before quantum theory, through Locke's concept of secondary qualities are those that arise from the relational interplay of objects and sense organs, but critics have tended to reject it in the hope of reaching primary (absolute) spatial and kinematic qualities.  It was only in the ambience of quantum mechanics that Heisenberg could proclaim the universality of the concept of secondary (interactional) quality – from that moment on, it became difficult (not to say impossible) to ignore that the entities and properties we pretend to “reveal” in the micro-domain cannot be dissociated from “subjective” agency.  

An essential feature of observation in quantum theory, which Heisenberg, Bohr, and Pauli all emphasized, is that there exist “complementary” pairs of physical properties – such as the position and momentum of a particle – that cannot be assumed to simultaneously possess determinate values, and the question of \textit{which} of the two properties may be assigned a determinate value depends on the experimental situation. In particular, it depends on the experimenter’s \textit{choice} of which apparatus to employ in performing the measurement, eg. in deciding whether to place a diffraction grating in front of an absorbing screen, enabling a measurement of a particle’s momentum rather than its position. By choosing (say) to insert the diffraction grating, the experimenter plays an active role in co-constituting the result, namely, a definite value of the momentum\footnote{Note that they do not “co-create” the result, as they do not have the power to determine which specific value of the momentum will be realized – cf. our earlier discussion concerning this distinction.
}. 

In this way, the experimenter’s \textit{agency}, through their ability to decide the experimental arrangement, becomes an integral component of the observational act. Classically the observer’s contribution to the secondary qualities resided only in their passive receptivity to external stimuli, leaving the door open to the possibility of removing the observer’s contribution altogether; the essential role played by the observer’s agency in quantum theory closes this door once and for all. Thus, the shift from classical to quantum physics resides not in the recognition that knowledge about physical systems is co-constituted (a fact already recognized classically) but rather in the recognition that such co-constitution is an \textit{ineliminable} feature of knowledge acquisition, which cannot be discounted as an obstacle that could be overcome in principle by improved methods and instrumentation.     

The ineliminable contribution of the observer’s agency in quantum theory only exacerbates the problem of objectivity that Kant sought to explicate and resolve in the context of classical physics. Following Kant, then, the first step towards a solution is to give up on the idea that objectivity seeks to attain knowledge of objects as they are independently from our agency, replacing it with a notion of objectivity that (i) refers to interactional phenomena and (ii) has “necessary universal validity” (Kant 2004). What is at stake, then, is not the wholesale removal of the observer’s contribution, but merely its universalization; that is, the task of objectivity is to establish a common meaning for the observation, which can be communicated between agents in a standardized shared language – allowing us to go beyond the view of a particular agent \textit{qua} individual “subject” to attain the view of an \textit{anonymous} member of a community of agents. 

Such a non-standard definition of objectivity, then, does not exclude subjectivity; instead it works to free itself from its too exclusively personal components and introduces us to a class of epistemologies in which even the subject/object divide must be considered problematic. Such a class of approaches in which objects are to be “constituted” as focal points of a multiplicity of subjective views was already considered by Kant and other philosophers, yet it is only in the context of quantum theory that this class of approaches appears to be \textit{indispensable} for understanding how objectivity can arise. 

\section{Representation—Map or Territory?}

We now turn to the second issue in which quantum mechanics reveals problems that already occur in classical physics.  “Representation,” much like the terms “subjective” and “objective,” is deeply embedded in the standard language used to describe all science. In common usage, “representation” is often treated as synonymous with “visual representation”. Diagrams, graphs, or images are considered prime examples because they are taken to \textit{stand for} the phenomena or entities under investigation, offering a more intuitive rendition of otherwise abstract or inaccessible domains. The same intuitive sense of representation holds for familiar scientific models such as the Bohr model of the atom or Watson’s and Crick’s cardboard model of DNA. Even in cases where there is widespread recognition that a model lacks this supposed correspondence to reality and veridicality—as with the simple Bohr model—the assumed function of representation is to render the invisible visible, the complex tractable, or the perplexing intelligible through perceptual means.

The nature of representation has also been one of the most hotly debated topics in mainstream analytic philosophy of science. Here, once again, we find questions that transcend the so-called quantum/classical limit.  While many philosophers agree that representation is central to understanding scientific theorizing and practice, the intuitive assumption that representation is, by default, visual (or imaginatively visual i.e. in the “minds eye”) has come under increasing scrutiny. 

Today, it is widely accepted that visual resemblance—long taken for granted in scientific pedagogy and communication—is neither a necessary nor a sufficient condition for scientific representation. For instance, highly abstract mathematical models or data sets may function as representations despite lacking any pictorial similarity to their targets, as in the case of evolutionary trees. Such trees do not depict the actual shapes or appearances of organisms but instead encode hypothesized relations of descent and divergence; their representational function lies in the structural and inferential roles they play within evolutionary biology rather than in visual similarity. Conversely, an image may closely resemble a phenomenon without being employed, within a scientific context, as a representation of it–such as artistic impressions of exoplanet discoveries. 

These striking depictions borrow the language of observation but are not the result of direct imaging or measurement; they are heuristic or aesthetic tools designed to evoke rather than to describe.  For these and related reasons, philosophical analysis in this area often begins with the formal assertion that “representation is a relation R such that the assertion that ‘X represents Y’ is equivalent to the assertion that ‘R holds between X and Y’” (Suárez 2010, p91). Based on this minimal understanding, the task then becomes one of characterizing the nature of R. 

Some do this in terms of similarity (Giere 2004, 2010), structural features such as isomorphism, partial isomorphism, or homology (Pero and Suárez, 2016; Da Costa and French, 2003; Ibarra and Mormann, 2005), or the inferential practices that R \textit{enables} (Suárez, 2010; for a general overview see Frigg and Nguyen 2020). Although the idea that the primary aim of scientific activity is to represent the world is widely accepted, there are, of course, dissenting views. Some, like Hacking (1983), argue for a general rejection of representation as a necessary component of scientific activity. Others take an anti-representationalist stance only within specific theoretical contexts, such as quantum physics (Fuchs, 2010; Healey, 2017).

Similarly to our earlier discussion of the objective/subjective divide, we begin our exploration of the notion of “representation” by introducing a basic distinction. If we understand representation broadly as a relation between X and Y, where X “stands in for” or “establishes reference to” Y, then all (non-formal, i.e. non-mathematical) theories are necessarily representational.  However, the triviality of this observation suggests that when we speak of scientific representation, we typically mean something more substantial. Specifically, for a theory or model to represent, it must not only establish reference to something beyond itself, but also imply that this “something” is part of a reality that exists independently of the theory or model—and independently of the agent devising and using it.

Clearly distinguishing between these senses of representation is crucial for recognizing that the prevailing, more substantial notion of representation is closely tied to the kind of dualism we discussed earlier in relation to subjectivity and objectivity. Starting from the dualist framework, which posits a strict separation between consciousness and the world, theories and models are said to enable conscious agents to interact with empirical target systems—albeit indirectly—through the former’s supposed representational capacities. Building on the well-known “no-miracles” argument for scientific realism (e.g. Leibniz’s pre-established harmony), the high predictive success of certain theories and models is then interpreted as evidence of “true” representation—that is, the theory’s or model’s capacity to accurately mirror the empirical target system it seeks to describe. However, if the properties and identity of the “target system” cannot be made independent of the project and instruments of its experimental probing, as is the case in quantum physics, then this standard hierarchy of predictive success and mirror-like description must be reversed. If no faithful picture of an “independent reality” can be taken at face value, then any representational mathematical structure becomes subordinate to its predictive function: it may still be useful for theorists to play with it, but only as a tool to facilitate the formulation of new predictions. To summarize, then: Operating within a dualist framework, rejecting straightforward realism leaves little room but to retreat into instrumentalism, since representation can no longer bridge the gulf it presupposes between theory and world.

Given our critique of the traditional dualism between the conscious observer and the world, it should come as no surprise that we reject the general framework within which the standard debate commonly unfolds.  Our rejection rests not only on philosophical grounds but, crucially, also on scientific considerations. One line of argument arises from discussions that were prevalent during the early days of quantum physics.  In this way we can see how this rejection forms the basis for re-examining the idea that classical physics is the low action limit of quantum physics.  

For Niels Bohr, for example, Planck’s discovery of the universal quantum of action $h$ captures the very essence of quantum theory because it challenges the conventional understanding of the relationship between observer and world. Crudely put, Bohr sees measurement (or any physical interaction) as involving an exchange of energy. This means that if we wish to know something about the measured system in itself, we must subtract the energy imparted to the system during the act of measurement. In classical physics, where energy is assumed to be infinitely divisible, this subtraction can be performed with arbitrary precision, making the distinction between observer and system seem unproblematic. However, if energy is quantized in units of $h$, this process encounters a fundamental limit. At the scale of $h$, the energy imparted cannot be subtracted out, and the boundary between observer and system becomes indeterminate. Because of the indivisibility of the quantum, we are, according to Bohr, “continually reminded of the difficulty to distinguish between subject and object” (Bohr 1934, p15).

This indivisibility has profound epistemological and metaphysical implications. Bohr argues that “objective description can be achieved only by including in the account of the phenomena explicit reference to the experimental conditions, [which] emphasizes in a novel manner the inseparability of knowledge and our possibilities of inquiry” (Bohr 1958, p12). Or, as Werner Heisenberg succinctly puts it, “What we observe is not nature in itself, but nature exposed to our method of questioning” (Heisenberg 1958, p58).

To many, the suggestion that knowledge can never—not even in principle—be separated from the activity of the observer, as well as from the methods and experimental techniques she employs, is profoundly radical because it challenges the very notion of scientific objectivity as the ongoing convergence toward a de-perspectivized view of reality. In response, quantum experience is treated either as a special, somewhat anomalous case, or efforts are made to develop interpretational frameworks that preserve classical conceptions of truth and objectivity, despite the persistence of quantum paradoxes. 

In this paper, we are arguing for the exact opposite approach: instead of treating quantum mechanics as confronting us with fringe cases that require special interpretational efforts, we propose that the true lesson of quantum mechanics is that the dualist assumption underlying more substantial forms of representationalism is fundamentally untenable for all parts of science. While it may be true that classical mechanics was founded on assumptions such as a clear separation between observer and system or the in-principle possibility of measurements that do not alter the state of the measured system, we view these assumptions as idealizations. Although they undoubtedly contributed to the remarkable success of classical science, they distort the fundamental relationship between conscious observers and reality. In fact, we see the notion that the conscious observer is merely a source of error to be eliminated for a truly scientific representation of any target system as yet another version of the blind spot—an echo of the old metaphysical lore that underpins much of our traditional thinking about science.

It is noteworthy that the mathematical foundations of quantum mechanics were extracted not from Newton’s version of mechanics but from the highly abstracted Hamiltonian version. Here the representation of systems occurs via coordinates in six dimensional (for a single particle) “phase space”. The dynamics is then analyzed in terms of the topologies of the energy surfaces (for energy conserving) systems.  Thus, even before the explicit introduction of the quantum, classical mechanics had moved from the simple representation of forces and corpuscles to the more purely mathematical representations that could not be directly visualized (no one can see a 6 dimensional object in their mind’s eye).  We note that this replacement of a formal system (a closed hyperdimensional surface) for the experienced one (particles in a box) represents what Edmund Husserl called the “surreptitious substitution”.  It is an extreme example of the dualist representational paradigm where the formal system, no matter how abstract, is taken to represent (or re-present) what appears to the conscious observer in a way that this representation is taken to be more real than what’s experienced, or at least have its own independent reality.  Thus, the problems arising with seeing the representation as independent of those carrying out the representation predate the rise of quantum physics.

\section{Implications for Physics}

Let us briefly summarize and refine our argument so that we can consider its implications. Quantum physicists have long recognized that quantum theory appears to challenge certain metaphysical ideas, which they took to be part-and-parcel of classical physics. Here we have focused on two such ideas: that physical systems and their properties have an existence independent of the processes by which we measure them; and that by making the results of our inquiry “objective” – that is, by removing or correcting for subjective contributions to the result – our physical theories give us insight into a reality that exists independently of our knowledge-gathering activities, viewed as it were “from nowhere”. While few would dispute that quantum theory challenges these notions, we have argued in the foregoing that, from a historical and philosophical perspective, there are many reasons to question the presumption that even \textit{classical physics} should be bound to these ideas.

What bearing does this have on contemporary discussions about the interpretation of Quantum Mechanics, and its relationship to classical physics? Physicists unwilling or unable to separate classical physics from the metaphysical ideas challenged by Quantum Mechanics are left with essentially two possible responses: either attempt to re-interpret Quantum Mechanics by trying to maintain its grounding into the purportedly classical metaphysical paradigm, or else declare that Quantum Mechanics is separated from classical physics by an insurmountable metaphysical rift. Within the latter category, one could posit a new metaphysics for quantum theory, emphasizing its dramatic departure from that of classical physics, or else one could reject metaphysical speculation altogether in favour of an instrumentalist or “anti-realist” philosophy. To provide some examples – not a complete survey: see Myrvold (2022) for a bibliography – the first option is exemplified in interpretations such as: the De Broglie-Bohm theory; psi-epistemic models; superdeterministic models; and more generally, all ontological models classified as hidden variable theories. The second option can be broadly associated with the “many-worlds” interpretation; the “Copenhagen school” of Bohr, Heisenberg, and Pauli; and many subsequent interpretations inspired by Copenhagen, such as Consistent Histories; and Relational Quantum Mechanics.

All these approaches, we argue, reside to varying degrees within the same blind spot: they assume that classical physics is necessarily bound to the aforementioned metaphysical commitments that Quantum Mechanics problematizes. There are, however, some noteworthy exceptions of interpretations that seek to restore continuity by revising the metaphysical premises not only of quantum theory but also classical physics. In the following we briefly discuss two such exceptions: the London-Bauer interpretation (1939), and QBism (circa 2009).

In 1939 physicists Fritz London and Edmond Bauer – hereafter “LB” – published a description of the quantum formalism which attempted to understand the problematic role of measurement (London and Bauer, 1939). For many years this work was mistakenly seen as supporting some version of a “consciousness causes collapse” argument like that explored by Wigner (French, 2023). Recently, however, French (2023) has demonstrated that LB were approaching quantum formalism from the perspective of Phenomenology; indeed, Fritz London had studied with colleagues of Edmund Husserl,the founding father of Phenomenology.   Phenomenology is well-known for its rejection of the kind of “naive realism” that we have here characterized as the idea that science gives us access to an independent reality that is cleanly separable from ourselves and our mode of inquiry; this perspective enabled LB to see that Quantum Mechanics did not merely break with preceding classical physical theories, but went further, undermining a set of far-reaching implicit metaphysical commitments. As LB write,

\begin{quote}
    In this way, the discussion of the [quantum] formalism taught us that the apparent philosophical point of departure of the theory, the idea of an observable world, totally independent of the observer, was a vacuous idea. Without intending to set up a theory of knowledge, although they were guided by a rather questionable philosophy, physicists were so to speak trapped in spite of themselves into discovering that the formalism of quantum mechanics already implies a well-defined theory of the relationship between the object and the observer, a relation quite different from that implicit in naive realism, which had seemed, until then, one of the indispensable foundation stones of every science.
\end{quote}

Thus we can argue that the question we raise about the relation between QM and CM as limits with regards to interpretation was recognized (at least by a few) early in the history of the field, and LB’s interpretation as described by French (2023) represents one of the earliest attempts to constructively go beyond the paradigm of “naive realism” – not only in Quantum Mechanics but in physics generally.
QBism, by comparison, provides an example of a well-developed contemporary interpretation of quantum mechanics that, like LB, seeks to interpret quantum theory in a way that goes beyond “naive realism”. Against the image of a ‘view from nowhere’, QBism proposes a “participatory universe” (a term coined by John Wheeler), in which “the agents (observers) matter as much as electrons and atoms in the construction of the actual world—the agents using quantum theory are not incidental to it [...] [Instead,] the agent is an active and non-negligible participator in the universe” (Fuchs, 2017a).  

In QBism, quantum states quantify the agent’s uncertainty about what they might experience in response to different measurements they might perform on a system. Despite some discursive vestiges of the old division between acting subjects and physical systems, QBism brings to the fore a concept that lies precisely at the place where subjectivity and objectivity meet and cannot be neatly divided into one side or the other, namely, the central concept of “agency” that is inherently “embodied” or “world-involved”.

This latter aspect of agency – embodiment – turns out to be crucial to how QBism challenges the subject-object divide. As we have seen, agency enters into Quantum Mechanics through the scientist’s role in choosing the particular experimental arrangement, and hence the physical property (say, position or momentum) that becomes manifest under measurement. Here, the material expression of agency is the \textit{apparatus}, itself usually taken to be a physical system, yet one that is not itself treated using the quantum formalism. The definition of the apparatus suffers however from a basic ambiguity that goes to the heart of the measurement problem: just which systems in a given experimental set-up should qualify as ‘apparatus’, and which ones as quantum systems? As Bohr and the Copenhagen school emphasized, if the quantum formalism is to be useful in any practical setting, then there must be a point at which the abstract vectors representing quantum states give way to results that are communicable in ‘common’ language – that is to say, the conceptual language of classical physics. But where to draw the line? 

For adherents to the Copenhagen interpretation, this ambiguity in itself represented a challenge to the subject-object divide (recalling Bohr’s ‘inseparability’ argument from Section 3). However, it remained just that: a conspicuous ambiguity, amounting to merely a \textit{negative} claim, asserting the impossibility of a perfect separation of subject from object. One of QBism’s central innovations, taking it beyond Bohr and the Copenhagenists, is its attempt to fill this lacuna with a \textit{positive} principle. 

QBism asserts that ‘a measurement apparatus must be understood as an extension of the agent himself, not something foreign and separate’, much like a prosthetic hand (Fuchs, 2017b). As Pienaar (2020) elaborates, this has the radical consequence that the division between agent/apparatus and quantum object is not fixed once and for all but is constantly being enacted in every measurement – in particular it changes to reflect dynamic alterations in the agent’s embodied capacities for perception. This account goes beyond merely problematizing the subject-object split: it grounds this distinction in the materially enacted boundary by which the agent delineates itself from the external world through ongoing interaction. Interestingly, this opens an unsuspected link between Quantum Mechanics and the philosophy of embodied agency and consciousness.
 
By grounding itself strictly in the embodied perspective of some agent, QBism also casts the quantum state in a new light, with repercussions for the notion of objectivity and the role of ‘representation’. According to QBism, quantum states do not describe an independent reality; rather they represent a guide to action, specific to the sensorimotor capabilities of the agent assigning the state. Formally, quantum states are re-cast as catalogs of probabilities, expressing the agent’s subjective expectations about what they will perceive (the outcomes) conditional on each possible choice of action (that is, on each possible choice of apparatus or ‘prosthetic extension’ they could use to make the measurement). 

By making both quantum states and the outcomes of measurements personal to each agent, QBism might seem to lack the resources needed to give a satisfactory account of scientific objectivity. As QBism explicitly takes the agent-world relation as a fundamental element, it easily sidesteps simplistic charge of solipsism; however, Steven French has challenged QBism to explain how even a basic notion of intersubjectivity – arguably a pre-requisite for objectivity – can come about (French, 2023).

QBism’s strategy in defending itself against this type of challenge has been to appeal to a ‘post-Kantian’ or more broadly Phenomenological notion of objectivity of the kind we discussed in Section 2. More specifically, ‘objectivity’ in QBism does not signify access to a world independently of agency, rather, it signifies an invariance across some – usually only implicitly defined – community of agents. As Pienaar (2025a) explains:

\begin{quote}
    QBism asserts that quantum theory is a ‘tool of reasoning’, that is, a set of norms or best practices that guide the decisions of agents. Implicit in this formulation is the idea that the users of quantum theory belong to a special class of agents, i.e. not only must they be capable of reasoning explicitly using symbolic language, but they must have specific concerns for which the usage of quantum theory confers an advantage, such as building a quantum computer, or designing high-precision particle detectors. [...] Given a specification of the community of possible agents, one may call “subjective” those quantities or entities which may vary from one agent to another within the community. Conversely, those entities which preserve their form across the whole community may be called “objective”, provided we understand this word as meaning “invariant between subjects” and not “independent of subjects”.
\end{quote}

Having established this, objectivity is then possible insofar as it is possible for agents to ‘share’ experiences (i.e. of measurement outcomes). As emphasized by Fuchs (Crease and Sares, 2021) and Pienaar (2025b), nothing in QBism precludes such sharing, provided certain conditions obtain – for instance, that one agent should not model the other as a superposed quantum system, as in the ‘Wigner’s friend’ thought experiment (for more discussion, see DeBrota et al., 2020, and Ch.7 of Pienaar, 2025b). Thus, far from excluding objectivity, QBism takes a constructive step in articulating the conditions that make objectivity possible.

Moreover, by casting Quantum Mechanics as a ‘tool of reasoning’, rather than as a description of reality, QBism rejects the strong notion of representation that we criticised in Section 3. Therefore, insofar as the formalism of Quantum Mechanics provides an insight into reality, it tells us something about the world as it is \textit{in relationship to agency}, not about the world as it is ‘in itself’ – the latter concept being incomprehensible in QBism. 

LB and QBism are therefore two examples where Quantum Mechanics has prompted physicists to construct alternatives to the “naive realist” metaphysical worldview; the fact that they are based on different philosophies and principles further indicates that a diversity of such views is possible.

We close with one final example that goes to the very root of the tension inherent in naive realist approaches: the evolution of the concept of a “true value” in \textit{metrology}. Defined as the “science of measurement and its application” by the Joint Committee for Guides in Metrology (JCGM) (2012), or more informally, the science of “measuring measurement” (Brown, 2021), metrology’s aims are clearly distinct from the physicist’s goal of (ostensibly) measuring nature – yet the two disciplines are inextricably connected. On one hand, the metrologist’s work of designing and realizing measurement standards, as well as calibrating instruments to these standards, depends crucially on detailed physical models; on the other hand, by underwriting the trustworthiness of measurement outcomes, metrology provides the foundation of all physicists' claims to knowledge about the world. 
As a supporting structure for scientific activity, when metrology performs its function well it tends to pass unnoticed by the very scientists who depend upon it (Brown, 2021). As an illustration of this particular ‘blind spot’, when in 1999 James Faller of NIST was invited by the American Physical Society (APS) to give a lecture about his work on precision measurements of gravity, he recounted his surprise at not finding “precision measurement” among the APS's supposedly comprehensive list of “ALL Areas of Physics” (Faller, 2000):

\begin{quote}
    My sense is that talking about physics without recognizing the role of measurement science is like trying to write a poem without meter or an organ piece without a pedal line. (On inquiring, I was told they “simply didn't have enough space.” I note, however, that the builders of most edifices do find space for a foundation.)
\end{quote}

The danger of taking metrology for granted is that we may forget that metrology itself is grounded on deep philosophical assumptions. Of special relevance is the concept of a \textit{true value} of a quantity, which has played both normative and conceptual roles in traditional measurement theory, especially within what later came to be known as the \textit{Error Approach}. As the third edition of the International Vocabulary of Metrology (VIM) explicitly states, the objective of measurement in this framework is to “determine an estimate of the true value that is as close as possible to that single true value” (JCGM, 2012). 

Rooted in the statistical frameworks of Gauss and Laplace, this conception treats the true value as an ideal quantity that measurement aims to access—even if it cannot itself be measured (Grégis, 2015). Implicit in this framework is a dual commitment: the true value is \textit{ontologically real}—existing whether or not it is measured—but \textit{epistemically inaccessible}. On this view, measurement error, expressed as a difference $E = x - v$ between the measured value $x$ and the true value $v$, is meaningful only if $v$ refers to a definite, well-defined quantity that exists independently of the measurement.

In recent decades, however, this view has been increasingly replaced by an \textit{Uncertainty Approach}, formalized in standards such as the Guide to the Expression of Uncertainty in Measurement (GUM) (JCGM, 2008), first published in 1995. The Uncertainty Approach treats measurement outcomes not as approximations of a hidden true value, but as probability distributions that express warranted belief given limited information. 

Through this reconceptualization of the notion of true value, metrologists have moved away from the traditional conception of measurement as aiming to approximate a hidden external reality, emphasizing instead the aim of attaining mutual coherence of measurement results obtained by diverse methods and instruments, together with the explicit modeling that goes into interpreting these results. \textit{Nowhere is the concept of objectivity more clearly shown to be a construction than in metrology, where the objectivity of results (in the sense of their independence across difference measurement contexts) is not pre-given, but must be painstakingly ensured through an explicit set of shared practices and methods.} 

Although these developments in metrology occurred long after the advent of quantum theory, it is significant that quantum theory played no role whatsoever in bringing about the downfall of the strong metaphysical realist reading of true values: rather, the foremost concerns already existed in classical metrology, and stemmed from the tension already inherent in presupposing that measurements can hope to reveal an external reality when, ultimately, the observed indications of measuring apparatuses can only ever be compared to observed indications of other apparatuses. This apparent circularity, referred to as \textit{the problem of nomic measurement} (Chang, 1995) turns out not to be a \textit{vicious} circle in the logician’s sense, as it still allows knowledge of the world to be possible; the caveat is that knowledge acquired under this condition cannot be divorced from the human activity of measuring; rather it requires making this activity an explicit component of physical knowledge.  

To conclude, the preceding examples from physics and metrology show how quantum theory does not represent a break with classical physics \textit{per se}, but rather brings to light and sharpens philosophical issues that were already latent in classical physics. These examples also serve as possible starting points for reconceptualizing the metaphysical framework underlying not just Quantum Mechanics, but physics in general.

\section{Implications beyond Physics}

We believe it is critical to note that, in our view, rejecting the usual story of classical mechanics, quantum mechanics and limits between them has more than theoretical implications.  Let us consider the dualist assumption of a strict separation between observer and system. Moving beyond the classical image of scientific representations as ideally de-humanized depictions of objective reality has also far-reaching \textit{existential} consequences. This was one of the central messages of the Austrian-German philosopher Edmund Husserl in his last major work, \textit{The Crisis of European Sciences}. 

Husserl’s argument, in a nutshell, is this: The undeniable success of modern science can foster an uncritical attitude in which we employ the mathematical apparatus of our most advanced theories without reflecting on the fact that this machinery did not descend from the proverbial sky but was constructed for specific human purposes. When we forget about the “life-world” roots of our theories, the temptation arises to project features of ideal models back onto nature itself, thus succumbing to the mistake to “take for \textit{true being} what is actually a \textit{method}” (Husserl 1970). This conflation between method and being is, for Husserl, the cardinal sin of the objectivist mindset. Instead of recognizing the constructive nature of models and the way in which they function as cognitive lenses for the constitution of the scientific image, we mistake these models and theories for truthful representations of the deep structure of reality. 

The result is a deep rift: the “world” represented by these models becomes increasingly irreconcilable with the world as we experience it in everyday life. Objectivism thus drives a wedge between life-world and science. Paradoxically, the life-world—the unsurpassable ground of all our scientific and pre-scientific practices—is relegated to the status of an illusion, while the “real world”—the world about which science speaks through its models—is rendered inaccessible to human experience. According to Husserl, it is precisely through this process that science begins to lose relevance for human existence: the world it speaks of becomes estranged from the world in which the joys, challenges, and tragedies of our lives actually unfold.
 
\section*{Funding}

This publication was made possible through the support of Grant 62424 from the John Templeton Foundation. The opinions expressed in this publication are those of the authors and do not necessarily reflect the views of the John Templeton Foundation. GBL also acknowledges support from FAPERJ, CIência Pioneira, CNPq, CAPES. JP acknowledges support from the Conselho Nacional de Desenvolvimento Científico e Tecnológico (CNPq) Grant/Process number PV351823/2025-5.

\section*{References}

\begin{HarvardBibliography}

Bitbol, M. (2002). ``Science as if Situation Mattered'', \textit{Phenomenology and the Cognitive Sciences}, 1, pp.181--224.\par

Bohr, N. (1920). ``Über die Serienspektra der Elemente'', \textit{Zeitschrift für Physik}, 2 (5), pp.423--478.\par

Bohr, N. (1934). \textit{Atomic Theory and the Description of Nature}. Cambridge: Cambridge University Press.\par

Bohr, N. (1958). ``The Unity of Human Knowledge'', in N. Bohr, \textit{Essays 1958--1962 on Atomic Physics and Human Knowledge}. New York: Wiley, pp.8--16.\par

Boltzmann, L. (1979). ``Über die Unentbehrlichkeit der Atomistik in der Naturwissenschaft'', in L. Boltzmann, \textit{Populäre Schriften}. Braunschweig: Vieweg, pp.78--93.\par

Brown, R.J.C. (2021) ``Measuring measurement - what is metrology and why does it matter?'' \textit{Measurement : journal of the International Measurement Confederation}, vol. 168, no. 108408.\par

Cabello, A. (2017). ``Interpretations of Quantum Theory: A Map of Madness'', in O. Lombardi, S. Fortin, F. Holik, and C. López (eds.), \textit{What Is Quantum Information}? Cambridge: Cambridge University Press, pp.138--144.\par

Chang, H. (1995). ``Circularity and Reliability in Measurement,'' \textit{Perspectives on Science}, vol. 3, no. 2, pp. 153--172.\par

Crease, R.P. and Sares, J. (2021). ``Interview with physicist Christopher Fuchs''. \textit{Continental Philosophy Review} 54 (4):541--561.\par

Da Costa, N., and French, S. (2003). \textit{Science and Partial Truth: A Unitary Approach to Models and Scientific Reasoning}. Oxford: Oxford University Press.\par

Daston, L. and Galison, P. (2007). \textit{Objectivity}. New York: Zone Books.\par

DeBrota, J. B., Fuchs, C., and Schack, R. (2020). ``Respecting One's Fellow: QBism's Analysis of Wigner's Friend''. \textit{Found Phys}. 50, pp.1859--1874.\par

Ehrenfest, P. (1927). ``Bemerkung über die angenäherte Gültigkeit der klassischen Mechanik innerhalb der Quantenmechanik'', \textit{Zeitschrift für Physik}. 45, pp.455--457.\par

Faller, J. (2000). ``Precision measurement with gravity''. \textit{Gravitational Waves and Experimental Gravity}, series: Proceedings of the XXXIVth Moriond Meeting, 1999, Hanoi, VN.\par

Frank, A., Gleiser, M. and Thompson, E. (2024). \textit{The Blind Spot: Why Science Cannot Ignore Human Experience}. Cambridge, MA: The MIT Press.\par

French,S. (2023). \textit{A Phenomenological Approach to Quantum Mechanics: Cutting the Chain of Correlations}. Oxford: Oxford University Press.\par

Frigg, R. and Nguyen, J. (2020). \textit{Modelling Nature: An Opinionated Introduction to Scientific Representation}. Cham: Springer.\par

Fuchs, C.A. (2010). ``QBism, the Perimeter of Quantum Bayesianism''. arXiv:1003.5209 [quant-ph].\par

Fuchs, C.A. (2017a). ``On Participatory Realism.'' In I. Durham and D. Rickles (eds.), \textit{Information and Interaction: Eddington, Wheeler, and the Limits of Knowledge}. Berlin: Springer.\par

Fuchs, C.A. (2017b). ``Notwithstanding Bohr, the Reasons for QBism.'' \textit{Mind and Matter} 15(2), pp.245--300.\par

Fuchs, C.A. (2023). ``QBism, Where Next?'', in P. Berghofer and H.A. Wiltsche (eds.), \textit{Phenomenology and QBism: New Approaches to Quantum Mechanics}. New York: Routledge, pp.78--143.\par

Ghirardi, G., and Bassi, A., (2020). ``Collapse Theories'' In E.N. Zalta and U. Nodelman (eds.), \textit{The Stanford Encyclopedia of Philosophy}. Summer 2020 Edition.\par

Giere, R.N. (2004). ``How Models Are Used to Represent Reality''. \textit{Philosophy of Science}, 71 (5), pp.742--752.\par

Giere, R.N. (2010). ``An Agent-Based Conception of Models and Scientific Representation''. \textit{Synthese}, 172 (2), pp.269--281.\par

Grégis, F. (2015). ``Can we dispense with the notion of `true value' in metrology?'' In O. Schlaudt, L. Huber (eds.), \textit{Standardization in measurement: philosophical, historical and sociological issues, History and philosophy of technoscience}. Pickering and Chatto, London. pp. 81 -- 93.\par

Hacking, I. (1983). \textit{Representing and Intervening: Introductory Topics in the Philosophy of Natural Science}. Cambridge: Cambridge University Press.\par

Harding, S. (1995). ``Strong Objectivity: A Response to the New Objectivity Question'', \textit{Synthese} 104 (3):331 - 349.\par

Healey, R. (2017). \textit{The Quantum Revolution in Philosophy}. Oxford: Oxford University Press.\par

Heisenberg, W. (1958). ``The Copenhagen Interpretation of Quantum Theory'', in \textit{Physics and Philosophy: The Revolution in Modern Science}. New York: Harper \& Brothers, pp.44--58.\par

Heisenberg, W. (1971). ``Positivism, Metaphysics, and Religion'', in \textit{Physics and Beyond: Encounters and Conversations}. New York: Harper \& Row, pp.205--217.\par

Ibarra, A. and Mormann, T. (2005). ``Interactive Representations''. \textit{Representaciones}, 1 (1), pp.1--20.\par

JCGM, (2008). \textit{Evaluation of measurement data --- guide to the expression of uncertainty in measurement}. JCGM 100:2008\par

Joint Committee for Guides in Metrology (JCGM), (2012). \textit{International vocabulary of metrology: basic and general concepts and associated terms (VIM)}. JCGM 200:2012, 3rd edition.\par

Kant, I. (1819). \textit{Logic}. Translated from the German by J. Richardson. London: W. Simpkin and R. Marshall.\par

Kant, I. (2004). \textit{Prolegomena to Any Future Metaphysics: That Will Be Able to Come Forward as Science}. 2nd edn. Edited and translated by G. Hatfield. Cambridge: Cambridge University Press.\par

Landsman, N.P. (2007). ``Between Classical and Quantum'', in J. Butterfield and J. Earman (eds.), \textit{Handbook of the Philosophy of Science: Philosophy of Physics}. Amsterdam: Elsevier, pp.417--553.\par

London, F., and Bauer, E. (1939). \textit{La théorie de l'observation en mécanique quantique}. Paris: Hermann. English translation: ``The theory of observation in quantum mechanics''. In J.A. Wheeler and W.H. Zurek (eds.), \textit{Quantum Theory and Measurement}. Princeton: Princeton University Press, 1983, 217--259.\par

Mach, E. (1919). \textit{The Science of Mechanics: A Critical and Historical Account of its Development}. Chicago and New York: Open Court.\par

Mermin, D. (1989). ``What's Wrong with This Pillow?'', \textit{Physics Today}, 42 (4), pp.9--11.\par

Myrvold, W. (2022). ``Philosophical Issues in Quantum Theory'', In E.N. Zalta and U. Nodelman (eds.), \textit{The Stanford Encyclopedia of Philosophy}. Fall 2022 Edition.\par

Nagel, T. (1986). \textit{The View from Nowhere}. Oxford: Oxford University Press.\par

Palacios, P. (2024). ``Intertheory Relations in Physics'' In E.N. Zalta and U. Nodelman (eds.), \textit{The Stanford Encyclopedia of Philosophy}. Spring 2024 Edition.\par

Pero, F. and Suárez, M. (2016). ``Varieties of Misrepresentation and Homomorphism''. \textit{European Journal for Philosophy of Science}, 6 (1), pp.71--90.\par

Pienaar, J.L. (2020). ``Extending the Agent in QBism.'' \textit{Foundations of Physics}. 50: 1894--1920.\par

Pienaar, J.L. (2025a). ``Unobservable entities in QBism and phenomenology''. In \textit{P. Berghofer and H}. Wiltsche, Phenomenology and QBism: New Approaches to Quantum Mechanics. Routledge.\par

Pienaar, J.L. (2025b). ``French on London and Bauer, and QBism''. \textit{Studies in History and Philosophy of Science}. 114, 102084.\par

Pineda, C. and Prosen, T. (2007). ``Universal and Nonuniversal Level Statistics in a Chaotic Quantum Spin Chain''. \textit{Physical Review E}, 76 (6), article 061127.\par

Planck, M. (1906). \textit{Vorlesungen über die Theorie der Wärmestrahlung}. Leipzig: Verlag von Johann Ambrosius Barth.\par

Shimony A. (1993). ``Search for a worldview which can accommodate our knowledge of microphysics'', in \textit{The Search for a Naturalistic World View}. Cambridge: Cambridge University Press. Pp.62--76.\par

Stern, R. (1998), ``The turn against idealism: materialism, positivism, empiricism, naturalism In: Nineteenth-century philosophy'', in \textit{the Routledge Encyclopedia of Philosophy Online}. Taylor and Francis. (Accessed online 1 April 2026).\par

Suárez, M. (2010). ``Scientific Representation''. \textit{Philosophy Compass}, 5 (1), pp.91--101.\par

Toole, B. (2022). ``Objectivity in Feminist Epistemology''. \textit{Philosophy Compass} 17 (11).\par

Vaidman, L. (2024), ``The Many-Worlds Interpretation of Quantum Mechanics: Current Status and Relation to Other Interpretations''. \textit{Quantum Reports} 6, no. 2: 142--146.\par

Weyl, H. 1949. \textit{Philosophy of Mathematics and Natural Science}. Princeton: Princeton University Press.\par

Wiltsche, H.A. (2025). ``The Coordination Problem: A Challenge for Transcendental Phenomenology of Science'', in B.C. Hopkins and D. De Santis (eds.), \textit{The New Yearbook for Phenomenology and Phenomenological Philosophy}. London: Routledge, pp.257--277.\par

\end{HarvardBibliography}

\end{document}